# A REVIEW OF TEV SCALE LEPTON-HADRON AND PHOTON-HADRON COLLIDERS

S. Sultansoy, Gazi University, Ankara, TURKEY

*Abstract*

The investigation of lepton-hadron and photon-hadron collisions at TeV scale is crucial both to clarify the strong interaction dynamics from nuclei to quark-parton level and for adequate interpretation of experimental data from future hadron colliders (LHC and VLHC). In this presentation different TeV scale lepton-hadron and photon-hadron collider proposals (such as THERA, "LEP"-LHC, QCD Explorer etc) are discussed. The advantages of linac-ring type colliders has been shown comparatively.

## INTRODUCTION

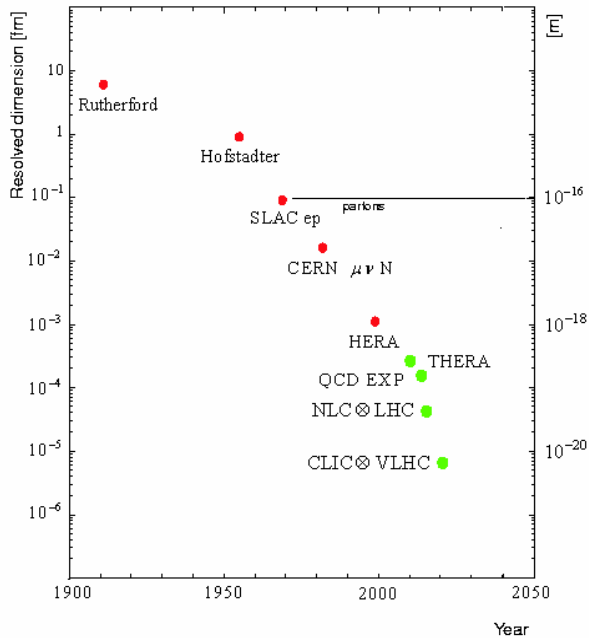

Figure 1: The development of the resolution power of the experiments exploring the inner structure of matter over time from Rutherford experiment to CLIC⊗VLHC.

It is known that lepton-hadron collisions have been playing a crucial role in exploration of deep inside of matter. For example, the quark-parton model was originated from investigation of electron-nucleon scattering. The HERA with $\sqrt{s} \approx 0.3$ TeV has opened a new era in this field extending the kinematics region by two orders both in high $Q^2$ and small x with respect to fixed target experiments. However, the region of sufficiently small x ($\leq 10^{-5}$) and simultaneously high $Q^2$ ($\geq 10$ GeV$^2$), where saturation of parton densities should manifest itself, is not currently achievable. The investigation of physics phenomena at extreme small x but sufficiently high $Q^2$ is very important for understanding the nature of strong interactions at all levels from nucleus to partons.

At the same time, the results from lepton-hadron colliders are necessary for adequate interpretation of physics at future hadron colliders. Concerning LHC, which hopefully will start in 2007, a $\sqrt{s} \approx 1$ TeV ep collider will be very useful in earlier 2010's when precision era at LHC will begin.

Finally, multi-TeV center of mass energy ep colliders are competitive to future hadron and lepton colliders in search for the BSM physics.

## TEV SCALE LEPTON-HADRON COLLIDERS

Today, linac-ring type machines seem to be the main way to TeV scale in lepton-hadron collisions (see [1, 2] and references therein). Construction of future linear collider or a special e-linac tangentially to existing (HERA, Tevatron, RHIC) or planned (LHC, VLHC) hadron rings will provide a number of new powerful tools in addition to ep and eA options:

- TeV scale γp [3] (see also [4]) and γA [5] colliders
- FEL-Nucleus colliders [6] (see also [7]).

### Standard Type ep Colliders

There are several standard (ring-ring) type ep collider proposals with $\sqrt{s} \geq 1$ TeV. The first one is an ep option for LHC, which assumes a construction of 67.3 GeV electron ring in the LHC tunnel [8]. Concerning the VLHC based ep collider, a construction of 180 GeV e-ring in the VLHC tunnel is proposed in [9]. However, a construction of an additional e-ring in the LHC and VLHC tunnels might cause a lot of technical problems (an example is inevitable removing of the LEP from the tunnel in order to assemble the LHC). Recently, linac-ring analogues of these proposals are discussed in [10]. It is shown that linacs will give opportunity to obtain the same $\sqrt{s}$ and luminosities with much shorter lengths.

Table 1: LHC and VLHC based ep colliders: e-ring vs e-linac (for TESLA-like linac)

| Collider | eLHC | eVLHC |
|---|---|---|
| $E_e$ (GeV) | 67.3 | 180 |
| $E_p$ (TeV) | 7 | 50 |
| $\sqrt{s}$ (TeV) | 1.37 | 6 |
| Ring circumference (km) | 26.66 | 531 |
| Luminosity ($10^{32}$cm$^{-2}$s$^{-1}$) | 1.2 | 1.4 |
| Linac length | 2.9 | 7.7 |
| Luminosity ($10^{32}$cm$^{-2}$s$^{-1}$) | 1.6 | 2.3 |

### THERA, ILC-Tevatron and QCD Explorer

Three versions of TESLA-HERA based ep collisions are considered in the TESLA TDR [11]: $E_e = 250$ GeV and $E_p = 1$ TeV with $L = 0.4 \times 10^{31} cm^{-2} s^{-1}$, $E_e = E_p = 500$ GeV with $L = 2.5 \times 10^{31} cm^{-2} s^{-1}$ and $E_e = E_p = 800$ GeV with $L = 1.6 \times 10^{31} cm^{-2} s^{-1}$.

Main parameters of ILC-Tevatron based lepton-hadron colliders are discussed in [12]. With nominal Tevatron parameters, the luminosity for $ep$ ($e\bar{p}$) collisions is calculated to be $8 \cdot 10^{29}$ $cm^{-2}s^{-1}$ ($4.6 \cdot 10^{29}$ $cm^{-2}s^{-1}$). The THERA [10] like upgrade of the proton beam parameters (namely, $\sigma_p = 10$ µm with $\beta_p = 10$ cm) leads to $L_{ep} = 1.2 \cdot 10^{31}$ $cm^{-2}s^{-1}$.

QCD Explorer assumes a collision of 75 GeV CLIC electron bunches with 7 TeV LHC proton beam [13, 14]. Super-bunch upgrade of the LHC will give opportunity to achieve $L_{ep} = 1.1 \cdot 10^{31}$ $cm^{-2}s^{-1}$ [13]. Otherwise, a radical upgrade of CLIC beam is necessary to achieve sufficiently high luminosity [10].

In spite of approximately equal center of mass energies, QCD Explorer is more advantageous than THERA and ILC-Tevatron for exploration of small $x_g$ region [10].

### "ILC"-LHC

The center of mass energy which will be achieved at this machine (0.5 TeV "ILC" electron beam on 7 TeV energy LHC proton beam) is an order higher larger than HERA. Certainly, $L_{ep} \approx 10^{32}$ $cm^{-2}s^{-1}$ is quite realistic estimation for "TESLA"-LHC (the factor 7 comparing to THERA is straightforward due to larger value of $\gamma_p$ at LHC). For "CLIC"-LHC, $L_{ep} \approx 10^{31}$ $cm^{-2}s^{-1}$ can be achieved with super bunch structure of LHC and nominal parameters of 0.5 TeV CLIC. The ep option, which extend both the $Q^2$-range and x-range by more than two orders of magnitude comparing to those explored by HERA, has a strong potential for both SM and BSM research. Concerning $\gamma p$ option, the advantage in spectrum of back-scattered photons will clearly manifest itself in a search for different phenomena. Rough estimations [2] show that the total capacity of ep and $\gamma p$ options for BSM physics (SUSY, compositness etc) research essentially exceeds that of a 0.5 TeV linear collider. Discovery limits for different phenomena obtained by "simple" rescale of corresponding results from [15] are presented in Figure 2. Detailed study for exited electrons [16] confirms "fingertip" estimations given in the Figure.

In the case of LHC nucleus beam IBS effects in main ring are not crucial because of large value of $\gamma_A$. The main principal limitation for heavy nuclei coming from beam-beam tune shift may be weakened using flat beams at collision point. Rough estimations show that $L_{eA} \cdot A \approx 10^{31}$ $cm^{-2}s^{-1}$ can be achieved at least for light and medium nuclei. For $\gamma A$ option, limitations on luminosity due to beam-beam tune shift is removed in the scheme with deflection of electron beam after conversion [3] and sufficiently high luminosity can be achieved for heavy nuclei, too. Certainly, nuclei options of "ILC"-LHC will bring out great opportunities for QCD and nuclear physics research. For example, $\gamma A$ option will five an opportunity to investigate quark-gluon plasma at very high temperatures but relatively low nuclear density (according to VMD, proposed machine will be at the same time $\rho$-nucleus collider).

### "CLIC"-VLHC

Concerning high energy frontiers, even 1 TeV e-linac will provide $\sqrt{s_{ep}} = 20$ TeV, whereas 3 (5) TeV CLIC will give $\sqrt{s_{ep}} = 34$ (45) TeV. Taking in mind THERA estimations one can expect $L_{ep} \approx 10^{33}$ $cm^{-2}s^{-1}$ for "ILC"-VLHC, whereas $L_{ep} \approx 10^{32}$ $cm^{-2}s^{-1}$ is rather conservative estimation for "CLIC"-VLHC. Let me remind that $\gamma p$ option will provide almost the same center of mass energy and luminosity as ep option. Obviously, Linac-VLHC will give opportunity to investigate a lot of particle physics phenomena in a best manner.

Table 1: Energy Frontiers

| Colliders | Hadron | Lepton | Lepton-Hadron |
|---|---|---|---|
| 1990's | Tevatron | SLC/LEP | HERA |
| $\sqrt{s}$, TeV | 2 | $0.1/0.1 \rightarrow 0.2$ | 0.3 |
| L, $10^{31} cm^{-2}s^{-1}$ | 1 | 1/1 | 1 |
| 2010's | LHC | "NLC"(TESLA) | "NLC"-LHC |
| $\sqrt{s}$, TeV | 14 | $0.5 \rightarrow 1.0(0.8)$ | $3.7 \rightarrow 5.3(4.7)$ |
| L, $10^{31} cm^{-2}s^{-1}$ | $10^3$ | $10^3$ | $1 \div 10$ |
| 2020's | VLHC | CLIC | "CLIC"-VLHC |
| $\sqrt{s}$, TeV | 200 | 3 | 34 |
| L, $10^{31} cm^{-2}s^{-1}$ | $10^3$ | $10^3$ | $10 \div 100$ |

## CONCLUSION

The importance of linac-ring type ep colliders was emphasized by Professor B. Wiik at Europhysics HEP Conference in 1993 [17]. Following previous article [18], he argued TESLA type accelerator to be used as linac. The argument is still valid for LHC-based ep, $\gamma p$, eA and $\gamma A$ colliders. Concerning VLHC-based ep and $\gamma p$ colliders, CLIC type linear accelerator seems to be advantageous, since the energy of TESLA of reasonable size is less than 1 TeV.

## REFERENCES

[1] S. Sultansoy, "Linac-Ring Type Colliders: Second Way to TeV Scale", European Physical Society International Europhysics Conference on High Energy Physics, Aachen, July 2003; Eur. Phys. J. C33 (2004) s1064.

[2] S. Sultansoy, "Four Ways to TeV Scale", First International Workshop on Linac-Ring Type ep and $\gamma p$ Colliders, Ankara, April 1997; Turkish J. Phys. 22 (1998) 575; http://journals.tubitak.gov.tr/physics/

[3] A.K. Ciftci et al., Nucl. Instr. Meth. A 365 (1995) 317.

[4] D. Schulte et al., "Photon-Nucleon Collider Based on LHC and CLIC", these Proceedings.

[5] A.K. Ciftci, S. Sultansoy and O. Yavas, Nucl. Instr. Meth. A 472 (2001) 72.

[6] H. Aktas et al., Nucl. Instr. Meth. A 428 (1999) 271.

[7] H. Braun et al., "CLIC Drive Beam and LHC Based FEL-Nucleus Collider", these Proceedings.

[8] E. Keil, LHC Project Report 93, CERN (1997).

[9] J. Norem and T. Sen, FERMILAB-Pub-99/347 (1999).

[10] L. Gladilin et al., hep-ex/0504008 (2005).

[11] H. Abramovitz et al., in TESLA TDR, DESY-2001-011, ECFA-2001-209 (2001); The THERA Book, DESY-LC-REV-2001-062; www.ifh.de/thera.

[12] O. Cakir et al., "Main Parameters of ILC-Tevatron based Lepton-Hadron Colliders", these Proceedings.

[13] D. Schulte and F. Zimmermann, "QCD Explorer Based on LHC and CLIC-1", EPAC'04, Lucerne, July 2004, p. 622, http://www.jacow.org.

[14] Mini-Workshop on Machine and Physics Aspects of CLIC Based Future Collider Options, CLIC Note 613, Geneva, 2004.

[15] U. Amaldi, Workshop on Physics at Future Accelerators, CERN Yellow report 87-07, p. 323 (1987).

[16] O. Cakir, A. Yilmaz and S. Sultansoy, Phys. Rev. D 70 (2004) 075011.

[17] B. Wiik, "Recent Developments in Accelerators", Europhysics HEP Conference, July 1993, Marseille.

[18] M. Tigner, B. Wiik and F. Willeke, Proc. of the 1991 IEEE Particle Accelerator Conference, p. 2910.

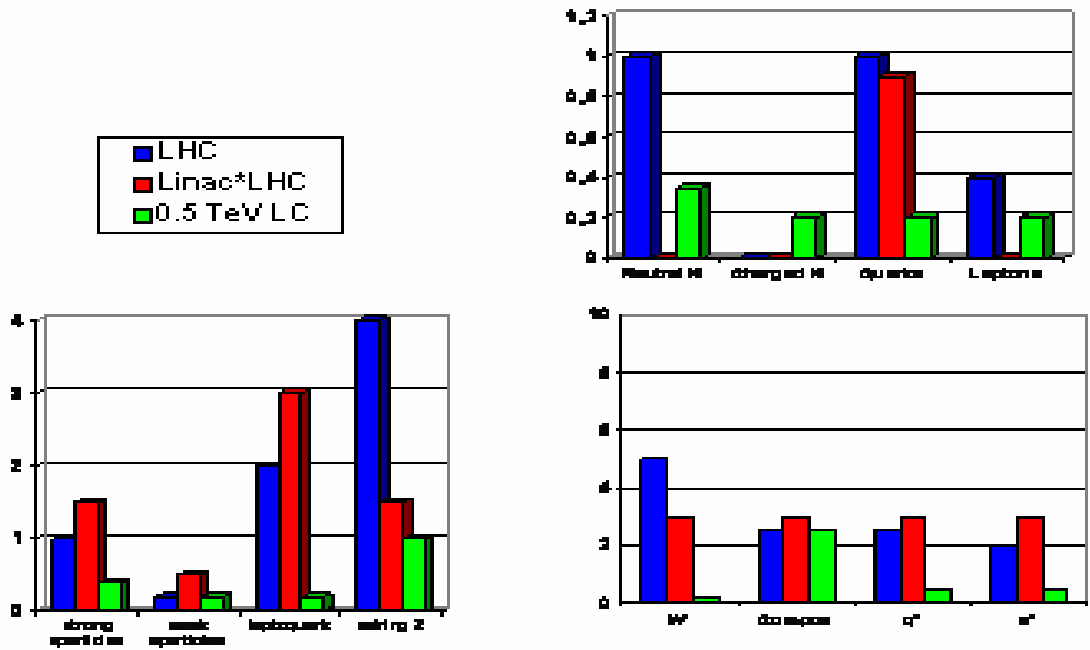

Figure 2: "Fingertip" estimations of discovery limits at the LHC (blue), ILC*LHC (red) and ILC (blue). Upper-left picture contains: the neutral Higgs, a charged Higgs, the fourth SM family quarks and leptons. Down-right picture contains: strong sparticles (gluino and squarks), weak sparticles (neutralino, chargino and sleptons), leptoquark and Z' from $E_6$. Down-left picture contains: W', compositness scale, excited quarks and leptons.